\documentclass[english,showkeys,showpacs,longbibliography,prb,twocolumn]{revtex4-2}
\usepackage[T1]{fontenc}
\usepackage[latin9]{inputenc}
\usepackage{babel}
\usepackage{amsmath}
\usepackage{amssymb}
\usepackage{graphicx}
\usepackage{color}
\usepackage[unicode=true,pdfusetitle,
 bookmarks=true,bookmarksnumbered=false,bookmarksopen=false,
 breaklinks=false,pdfborder={0 0 1},backref=false,colorlinks=false]
 {hyperref}

\newcommand{\VEC}[1]{{\boldsymbol{ #1}}}
\newcommand{\qvec}{{\VEC{q}}}

\newcommand{\invcm}{cm$^{-1}$}
\newcommand{\invA}{\AA$^{-1}$}

\newcommand{\mos}{MoS$_2$}
\newcommand{\bise}{Bi$_2$Se$_3$}
\newcommand{\half}{\frac{1}{2}}
\newcommand{\Fig}{{Fig.}}
 
\newcommand{\smd}{{\text -}} 



\begin{document}
\title{
Complementary local-global approach for phonon mode connectivities
}
\author{Chee Kwan Gan}
\email{ganck@ihpc.a-star.edu.sg}
\author{Zhun-Yong Ong}

\affiliation{Institute of High Performance Computing, 1 Fusionopolis Way, \#16-16 Connexis 138632, Singapore}
\date{Nov 16, 2020}
\begin{abstract}
Sorting and assigning phonon branches (e.g., longitudinal acoustic) of phonon modes is important for characterizing the phonon bands of a crystal and the determination of phonon properties such as the Gr\"uneisan parameter and group velocity. 
To do this, the phonon band indices  (including the longitudinal and transverse acoustic) have to be assigned correctly to all phonon modes across a path or paths in the Brillouin zone.  
As our solution to this challenging problem, we propose a computationally efficient and robust two-stage hybrid method that combines two approaches with their own merits.  
The first is the perturbative approach in which we connect the modes using degenerate perturbation theory.  
In the second approach, we use numerical fitting based on least-squares fits to circumvent local connectivity errors at or near exact degenerate modes. 
The method can be easily generalized to other condensed matter problems involving Hermitian matrix operators such as electronic bands in tight-binding Hamiltonians or in a standard density-functional
calculation, and photonic bands in photonic crystals.
\end{abstract}
\maketitle

\section{Introduction}

Lattice dynamical studies\cite{Born1956-book,VanDeWalle02v74} are indispensable for a detailed understanding of the thermodynamics, phase stabilities, phase transitions, and thermal properties of materials\cite{Grimvall1999-book,Mujica03v75,Yang12v98,Gan19v31}.  
The methods used to calculate the properties related to phonons fall into two general categories with their respective advantages. 
The methods in the first category are versatile since they apply density-functional perturbation theory (DFPT)\cite{Baroni01v73,Gonze97v55a} to a unit cell to compute analytic energy
derivatives\cite{Giannozzi09v21,Gonze09v180}. 
The methods in the second category use small atomic displacements\cite{Parlinski97v78,Kresse95v32,Togo15v108,Liu14v16,Wang16v2,Gan21v259} to numerically compute the interatomic force constants from the induced forces based on the Hellman-Feynman theorem\cite{Feynman39v56}.  
Common to these two categories of methods is the problem of dealing with small changes in the dynamical matrices for the evaluation of phonon frequency derivatives with respect to strain or volume for the Gr\"uneisen parameters\cite{Mounet05v71,Gan19v31,Gan18v151} or with respect to wavevectors (or $\qvec$ vectors) for phonon group velocities. 
In this case, the perturbation method from quantum mechanics is extremely useful\cite{Wallace1972-book,Gan15v92,Lee17v96}. 

\begin{figure}
\begin{centering}
\includegraphics[width=8cm]{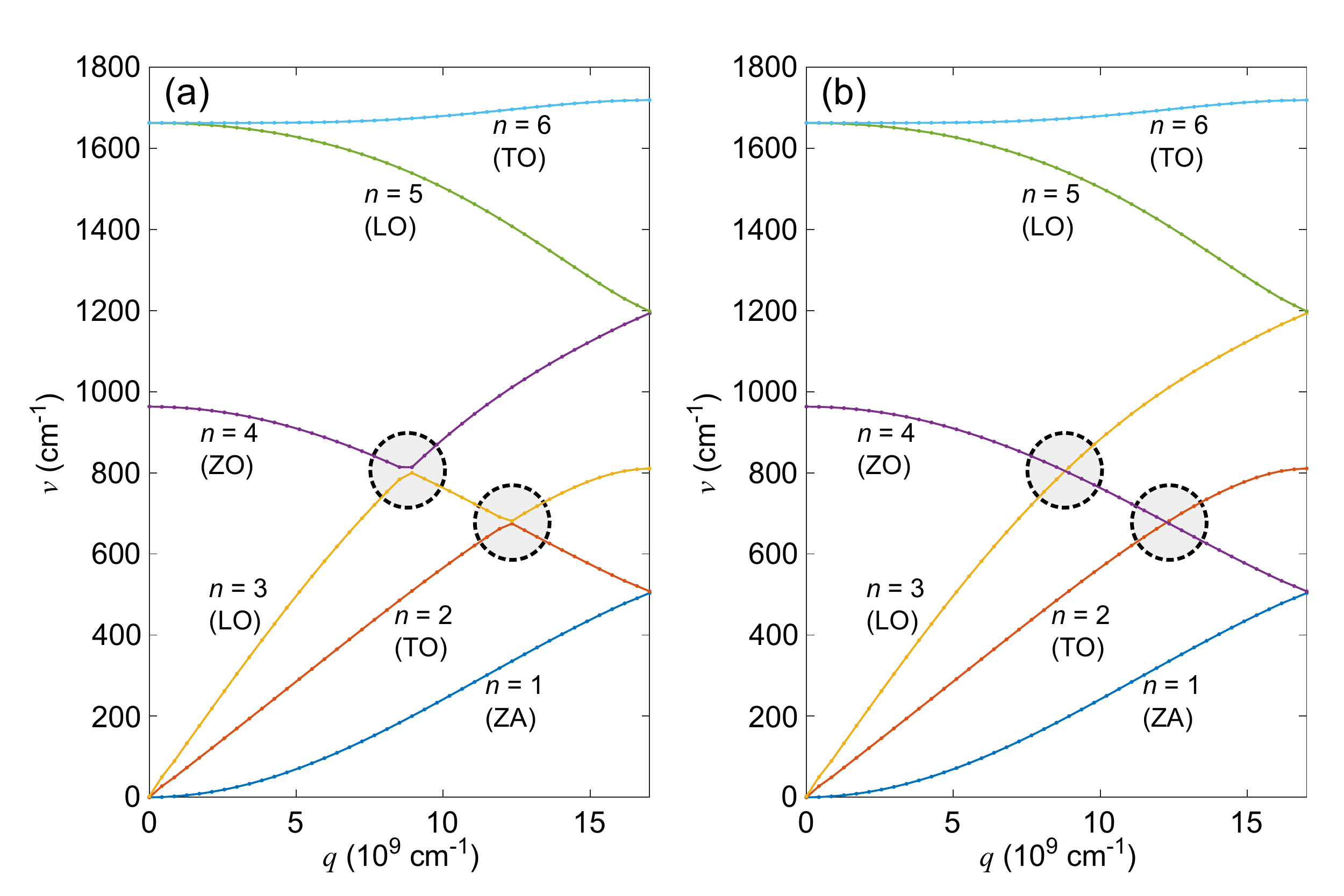}
\par\end{centering}
\caption{Phonon dispersion curves of graphene along the $\Gamma\smd{}K$ direction
with (a) frequency sorting and (b) physical phonon branch sorting.
Each phonon branch is labeled by a band index $n$ with its phonon
branch name in parenthesis. Each branch consists of 41 $\qvec$ points.
The crossing points are indicated by circles.}

\label{fig:UnsortedVsSortedPhononDispersion}
\end{figure}

However the problem of phonon mode connectivity is of a somewhat different nature than the calculation of frequency derivatives since the goal is to connect phonon modes (to form bands, analogous to the electronic counterpart\cite{Xiao10v82}) with neighboring $\qvec$ points and the connectivities are to be extended throughout the whole path in the Brillouin zone (BZ). 
When two neighboring phonon modes at $\qvec$ and $\qvec+\delta\qvec$ are connected, they are both assigned the same band index which groups them into the same phonon branch.
To connect two neighboring phonon modes, we need the eigenvalues and eigenvectors of the phonon modes which can be obtained by solving the eigenvalue problem\cite{MTDove:Book93_Introduction}
\begin{equation}
  D(\qvec)u_\qvec=\omega^2_\qvec u_\qvec
\label{eq:DynMatrix_eigenvalue}
\end{equation}
where $D(\qvec)$ is the dynamical matrix of dimension $N\times N$, and $u_\qvec$ and $\omega^2_\qvec$ are the associated eigenvector and eigenvalue, respectively. 
For a crystal, $N = 3n_c$ where $n_c$ is the number of atoms in the unit cell. 
When diagonalized, the Hermitian matrix $D(\qvec)$ in Eq.~(\ref{eq:DynMatrix_eigenvalue}) yields $N$ phonon modes with eigenvalues $\omega^2_{n,\qvec}$ (where $\omega_{n,\qvec}$ denotes the mode frequencies) and the corresponding eigenvectors $u_{n,\qvec}$, for phonon band index $n=1,2,\cdots, N$.
In the phonon mode connectivity problem, the phonon band indices should be assigned so that the band index of each phonon mode corresponds to the physical phonon branch to which it belongs. 
For instance, a phonon mode for graphene with $n=1$ should belong to the out-of-plane flexural acoustic branch (see Fig.~\ref{fig:UnsortedVsSortedPhononDispersion}).

In practice, the band index of a phonon mode is often assigned by ranking the frequencies of all phonon modes at the same $\qvec$ point such that 
$\omega_{1,\qvec}\leq\omega_{2,\qvec}\leq\cdots\leq\omega_{N,\qvec}$.
In a simple crystal, this assignment usually works when $\qvec$ is close to (but not at) the center of the BZ because the phonon modes sharing the same $n$ and $\qvec$ values usually belong to the same physical phonon branch at small $\qvec$. 
However, farther away from the BZ, this naive assignment of the band indices by phonon frequency ranking fails because the crossing of phonon branches changes the order of the eigenvalues, resulting in a breakdown of the association between phonon frequency ranking and the physical phonon branch. 
Around these crossing points, the use of phonon frequency ranking to connect the phonon modes at different $\qvec$ points results in illusory level repulsion~\citep{OYazyev:PRB02_Efficient}, as can be seen in Fig.~\ref{fig:UnsortedVsSortedPhononDispersion} which shows the graphene phonon dispersion curves along the $\Gamma\smd{}K$ direction with incorrect phonon frequency sorting and correct physical phonon branch sorting, with the phonon frequencies $\omega_{n,\qvec}$ obtained using the Tersoff interatomic potential~\citep{JTersoff:PRL88_Empirical,LLindsay:PRB10_Optimized}. 

One common solution to this problem is to utilize the similarities of eigenvectors between neighboring $\qvec$ points in the BZ~\citep{LHuang:PRB14_Correlation}.  
Suppose the band indices for the eigenmodes at $\qvec$ are assigned correctly with respect to the physical phonon branch. 
We can then assign the band indices for the eigenmodes of a neighboring point $\qvec+\delta\qvec$ through the eigenvector similarity condition
\begin{equation}
  |u_{m,\qvec+\delta\qvec}^{\dagger}u_{n,\qvec}|=\delta_{mn}+O(|\delta\qvec|)\ ,
  \label{eq:EigenvectorOverlapCondition}
\end{equation}
i.e., an eigenvector at $\qvec+\delta\qvec$ is assigned the band index $n$ if its overlap with an eigenvector at $\qvec$ with the band index $n$ is the closer to unity than that with any other eigenvectors at $\qvec+\delta\qvec$.  
The accuracy of the band index assignment in Eq.~(\ref{eq:EigenvectorOverlapCondition}) depends on the size of $\delta\qvec$ and may fail when $\delta\qvec$ is relatively large. 
The operational assumption behind Eq.~(\ref{eq:EigenvectorOverlapCondition}) is that under a small enough shift $\qvec\rightarrow\qvec+\delta\qvec$, the frequency $\omega_{n,\qvec}$ and eigenvector $u_{n,\qvec}$ change smoothly to $\omega_{n,\qvec+\delta\qvec}$ and $u_{n,\qvec+\delta\qvec}$, respectively. 
However, accurate assignment of the band index in Eq.~\ref{eq:EigenvectorOverlapCondition} requires the use of a very dense $\qvec$ grid in the BZ, which is computationally inefficient. 

Another problem with this eigenvector similarity approach is that it may fail at the crossing points that correspond to near or exact degenerate modes. 
The reason is that an arbitrary linear combination of eigenvectors of a degenerate eigenvalue is also an eigenvector with the same eigenvalue (to within a numerical tolerance). 
Therefore it is difficult to correctly connect modes based on the eigenvector similarity condition alone. 
One may bypass this problem by choosing a different $\delta \qvec $ during computing runtime to avoid the problematic $\qvec$ points although this entails a very complicated implementation where the outcome may still be questionable.
It is important to recognize that Eq.~(\ref{eq:EigenvectorOverlapCondition}) is entirely reliant on the local eigenvector similarity of neighboring $\qvec$ points which, near or at degeneracy points, becomes susceptible to assignment errors that can propagate across the BZ.
Hence, a more robust approach that circumvents local assignment errors is needed.

To solve this connectivity problem, we first discuss how the eigenvectors and eigenvalues of the dynamical matrices at neighboring $\qvec$ points can be related systematically 
within the framework of perturbation theory. 
Using this theoretical framework, we show how the overlap condition in Eq.~(\ref{eq:EigenvectorOverlapCondition}) can be derived and generalized.
Exploiting the perturbative approach for estimating eigenvalue, we propose a new two-stage hybrid method where in the first stage of the algorithm, an eigenvalue-perturbative approach is used to connect modes
of the neighboring $\qvec$ points
in the beginning of a path in the BZ.
However, this perturbative approach is error-prone near or at degenerate points as it only uses the local information around the $\qvec$ point.   
Therefore in the second stage of the algorithm, we switch to a more `global' approach that uses compatibility with the eigenvalues of the phonon modes connected in earlier stages as a criterion for assigning the band index. 
We call the second stage the fitting approach since it is based on the least-squares fits to the modes connected in the first stage.  
As we shall see, the hybrid method completely solves the phonon mode connectivity problem by using global information to circumvent local assignment errors.  

Our paper is organized as follows.  
In Section~\ref{sec:method}, we relate the eigenvectors and eigenvalues of the dynamical matrices at neighboring $\qvec$ points through the framework of perturbation theory,    
and show how the eigenvectors and eigenvalues can be connected locally in a systematic fashion.
The origin of the local band index assignment error is explained. 
We then present a least-squares fit approach to complement the eigenvalue-perturbative approach used for local phonon mode connectivities.  
In Section~\ref{sec:results}, we illustrate our two-stage hybrid method by presenting the results for three systems: graphene, \bise{}, and \mos{}.  
Finally we state our conclusions in Section~\ref{sec:conclusion}.

\section{Methodology}
\label{sec:method}

\subsection{Local mode connectivity}
\label{subsec:LocalConnectivity}

To explicate the relationship between the eigenvectors and eigenvalues at neighboring $\qvec$ points more systematically, it is necessary to recall that at $\qvec$, the $N\times N$ Hermitian dynamical matrix $D(\qvec)$~\citep{MTDove:Book93_Introduction}, and its associated eigenvalues $\omega^2_{n,\qvec}$ and eigenvectors $u_{n,\qvec}$, are governed by the equation
\begin{equation}
  D(\qvec)u_{n,\qvec}=\omega_{n,\qvec}^{2}u_{n,\qvec}\ ,
  \label{eq:DynMatrix}
\end{equation}
where $n$ denotes the phonon band index for $n=1,2,\cdots,N$ and $u_{n,\qvec}$ is the corresponding column eigenvector which satisfies the orthonormality condition $u_{m\qvec}^{\dagger}u_{n,\qvec =\delta_{mn}}$.
At $\qvec+\delta\qvec$, the corresponding dynamical matrix and its associated eigenvalues and eigenvectors are similarly given by 
\begin{equation}
  D(\qvec+\delta\qvec)u_{m,\qvec+\delta\qvec}=\omega_{m,\qvec+\delta\qvec}^{2}
  u_{m,\qvec+\delta\qvec}
  \label{eq:DynMatrix_neighbor}
\end{equation}
for $m=1,2,\cdots,N$.  
For orientational purposes, we assume that none of the eigenvalues in Eqs.~(\ref{eq:DynMatrix}) and (\ref{eq:DynMatrix_neighbor}) are degenerate, i.e., they do not form
crossing points. 
Traditionally, we assume that $u_{n,\qvec}$ and $u_{m,\qvec+\delta\qvec}$ belong to the same phonon branch if their overlap is close to unity, i.e.,
\begin{equation}
  |u_{m,\qvec+\delta\qvec}^{\dagger}u_{n,\qvec}|=\delta_{mn}+O(|\delta\qvec|)\ ,
  \label{eq:NaiveOverlapCondition}
\end{equation}
because under a small enough wavevector shift $\qvec\rightarrow\qvec+\delta\qvec$, the frequency $\omega_{n,\qvec}$ and eigenvector $u_{n,\qvec}$ deform \emph{smoothly} to $\omega_{n,\qvec+\delta\qvec}$ and $u_{n,\qvec+\delta\qvec}$, respectively, for most $\qvec$ points. 
The limitation of the eigenvector similarity condition in Eq.~(\ref{eq:NaiveOverlapCondition}) is that its accuracy diminishes rapidly as $|\delta\qvec|$ increases.

To relate Eqs.~(\ref{eq:DynMatrix}) and (\ref{eq:DynMatrix_neighbor}) formally, we treat $D(\qvec)$ and $D(\qvec+\delta\qvec)$ as the analogs of the unperturbed and perturbed Hamiltonian in perturbation theory, respectively. 
The analog of the perturbation term in this case is 
\begin{equation}
  V = D(\qvec+\delta\qvec)-D(\qvec) \ .	
  \label{eq:PerturbationTerm}
\end{equation}
To facilitate an expansion in a dummy variable $\lambda$, we consider a general perturbation term $\lambda V$ so that $u_{n,\qvec+\delta\qvec}$ is expanded as a power series in
$\lambda$, i.e., 
\begin{equation}
  u_{n,\qvec+\delta\qvec}
  =
  \sum_{j=0}^{\infty}\lambda^{j}u^{(j)}_{n,\qvec+\delta\qvec}\label{eq:EigModePowerSeries}
  \end{equation}
where the zeroth-order term ($j=0$) in the series is $u^{(0)}_{n,\qvec+\delta\qvec}=u_{n,\qvec}$.
Likewise, the eigenvalue $\omega_{n,\qvec+\delta\qvec}^{2}$ can be expanded as
\begin{equation}
  \omega_{n,\qvec+\delta\qvec}^{2}
  =
  \sum_{j=0}^{\infty}\lambda^{j}(\omega^{(j)}_{n,\qvec+\delta\qvec})^2 \label{eq:EigFreqPowerSeries}  
\end{equation}
where $\omega^{(0)}_{n,\qvec+\delta\qvec} = \omega_{n,\qvec}$.

The higher-order terms in the power series in Eq.~(\ref{eq:EigModePowerSeries}) can be obtained using the familiar nondegenerate perturbation theory in quantum mechanics. For instance, the first-order in Eq.~(\ref{eq:EigModePowerSeries}) ($j=1$) term is 
\begin{align}
u^{(1)}_{n,\qvec+\delta\qvec} 
 & =\sum_{m\neq n}\frac{u_{m,\qvec}^{\dagger}Vu_{n,\qvec}}{\omega_{n,\qvec}^{2}-\omega_{m,\qvec}^{2}}u_{m,\qvec}\nonumber \\
 & =\sum_{m\neq n}\frac{u_{m,\qvec}^{\dagger}D(\qvec+\delta\qvec)u_{n,\qvec}}{\omega_{n,\qvec}^{2}-\omega_{m,\qvec}^{2}}u_{m,\qvec} \ .
 \label{eq:EigModeFirstOrderCorrection}
\end{align}
The last equality follows because $u_{m\qvec}^{\dagger}D(\qvec)u_{n,\qvec}=0$ for $m\neq n$. Therefore, to the first order, we obtain 
\begin{equation}
  u_{n,\qvec+\delta\qvec} = u_{n,\qvec}+\lambda u^{(1)}_{n,\qvec+\delta\qvec}+O(\lambda^{2}) \ .
  \label{eq:EigModePowerSeries_expanded}
\end{equation}
Similarly, the first-order term in Eq.~(\ref{eq:EigFreqPowerSeries}) is
\begin{equation}
  (\omega^{(1)}_{n,\qvec+\delta\qvec})^{2} = u_{n,\qvec}^{\dagger}Vu_{n,\qvec} \ .
  \label{eq:FirstOrderFreqShift}
\end{equation}
If we set $\lambda=1$, then we can define the estimated eigenvector with band index $n$ at $\qvec+\delta\qvec$ as
\begin{equation}
  u^{\prime}_{n,\qvec+\delta\qvec} = u_{n,\qvec}+u^{(1)}_{n,\qvec+\delta\qvec} 
  \label{eq:EigenvecEstimator}
\end{equation}
and its associated eigenfrequency as
\begin{equation}
  \Omega^{2}_{n,\qvec+\delta\qvec} = \omega_{n,\qvec}^{2} + (\omega^{(1)}_{n,\qvec+\delta\qvec})^{2}\ . 
  \label{eq:EigenfreqEstimator}
\end{equation}

There are two ways that perturbation theory can be used for phonon mode connectivity given Eqs.~(\ref{eq:EigenvecEstimator}) and (\ref{eq:EigenfreqEstimator}).
The first way is to use Eq.~(\ref{eq:EigenvecEstimator}) to sort the eigenmodes at $\qvec+\delta\qvec$ with respect to the eigenmodes at $\qvec$ using a generalized version of Eq.~(\ref{eq:EigenvectorOverlapCondition}).
Suppose we have a set of eigenmodes at $\qvec$ with the correct band indices ($n=1,\cdots,N$) and their eigenvectors given by $u_{n,\qvec}$. 
If we have an eigenmode with eigenvector $u_{m,\qvec+\delta\qvec}$ satisfying Eq.~\ref{eq:DynMatrix_neighbor},
then we can assign it the band index $n$ if it satisfies the overlap condition 
\begin{equation}
  |u_{m,\qvec+\delta\qvec}^{\dagger}u^{\prime}_{n,\qvec+\delta\qvec}| = \delta_{mn}+O(\lambda^{2})\ .
  \label{eq:EigenmodeOverlap}
\end{equation}
where $u^{\prime}_{n,\qvec+\delta\qvec}$ is as defined in Eq.~(\ref{eq:EigenvecEstimator}). 
If we use only the first term for $u^{\prime}_{n,\qvec+\delta\qvec}$ on the right hand side of Eq.~(\ref{eq:EigenvecEstimator}) and substitute it to Eq.~\ref{eq:EigenmodeOverlap}, we recover Eq.~(\ref{eq:EigenvectorOverlapCondition}). 
The accuracy of the overlap condition of Eq.~(\ref{eq:EigenmodeOverlap}) may be increased by using a higher-order estimate for Eq.~(\ref{eq:EigenvecEstimator}) at the cost of increasing complexity in implementation.

In the second way, we connect the eigenmodes at $\qvec$ and $\qvec+\delta\qvec$ by comparing the estimated eigenvalues $\Omega_{n,\qvec+\delta\qvec}^{2}$ from Eq.~(\ref{eq:EigenfreqEstimator}) to the exact eigenvalues $\omega_{m,\qvec+\delta\qvec}^{2}$ from Eq.~(\ref{eq:DynMatrix_neighbor}).
We assigned the band index of $n$ to $\omega_{m,\qvec+\delta\qvec}^{2}$ (i.e., $m=n$) if the difference between $\Omega_{n,\qvec+\delta\qvec}^{2}$ and $\omega_{m,\qvec+\delta\qvec}^{2}$ is minimum.

We now provide more details on the implementation of the second way mentioned above.
\footnote{See an implementation of our algorithm in the subroutine `bandconnect' in qmod.f90 from https://github.com/qphonon/band-connectivity}
From two chosen points $\qvec_b$ and $\qvec_e$ that specify a path along a high symmetry direction in the BZ, we may define $(p+1)$ $\qvec$ points
\begin{equation}
  \qvec_i = \qvec_b + i(\qvec_e- \qvec_b)/p, \ \ i = 0, 1, \cdots, p 
\end{equation}
to uniformly connect $\qvec_b$ to $\qvec_e$.  
The $(p+1)$ dynamical matrices $D(\qvec_i)$ at $\qvec_i$ are evaluated and diagonalized to obtain $(p+1)$ lists of increasing eigenvalues $\omega^{2}_{n,\qvec_i}$, where $n=1,2,\cdots, N$.  
Due to possible band crossings, we cannot simply join the values of $\omega_{n,\qvec_i}$ for a particular $n$ value, for $i=0,1,\cdots, p$ and identify them as the $n$th phonon band.

To connect the frequencies at $\qvec_i$ to the frequencies at $\qvec_{i+1}$ (obviously $i=0$ for a fresh start), we use the following method.  
At $\qvec_i$, we have the phonon frequencies $\omega_{n,\qvec_i}$ and their associated eigenvectors $u_{n,\qvec_i}$, $n=1, 2, \cdots, N$. 
From these eigenvalues and eigenvectors, we find the perturbed (or predicted) eigenvalues $\Omega^2_{n,\qvec_{i+1}}$ at $\qvec_{i+1}$, correct up to the first order, through $\Omega^2_{n,\qvec_{i+1}} = \omega^{2}_{n,\qvec_i} + \Delta_{n,\qvec_i}$ where the correction terms $\Delta_{n,\qvec_i}$ are to be deduced from the first-order perturbation theory with the perturbation term $V_i = D(\qvec_{i+1}) - D(\qvec_i)$ as in Eq.~(\ref{eq:PerturbationTerm}).  

However, due to the presence of possible degenerate eigenvalues at $\qvec_i$, we must apply \emph{degenerate} first-order perturbation theory to find the correction terms $\Delta_{n,\qvec_i}$.  
Specifically this is achieved by first partitioning all $N$ eigenvalues $\omega^2_{n,\qvec_i}$ into a number of clusters of eigenvalues, where each cluster consists of $d$ eigenvalues (where in general two clusters may have different $d$) such that the frequency of any mode in the cluster is close to the frequency of at least one other mode in the cluster by a tolerance $\tau$. 
Hence, each cluster represents a numerically degenerate subspace for which the first-order correction has to be computed separately. 
We obtain for each cluster a $d \times d$ matrix $A$ with matrix elements $A_{\alpha\beta} = u_{\alpha,\qvec_i}^\dagger V_{i} u_{\beta,\qvec_i}$. 
A diagonalization of $A$ gives $d$ first-order corrections $\Delta_{n,\qvec_i}$ for $d$ eigenvalues $\omega^2_{n,\qvec_i}$ in the cluster.  
When a mode is singly degenerate (or nondegenerate, i.e., $d=1$), a cluster consists of just one distinct eigenvalue and we recover the result of the standard
nondegenerate first-order perturbation theory in Eq.~(\ref{eq:FirstOrderFreqShift}).

\begin{figure}
\centering\includegraphics[width=9.2cm,clip]{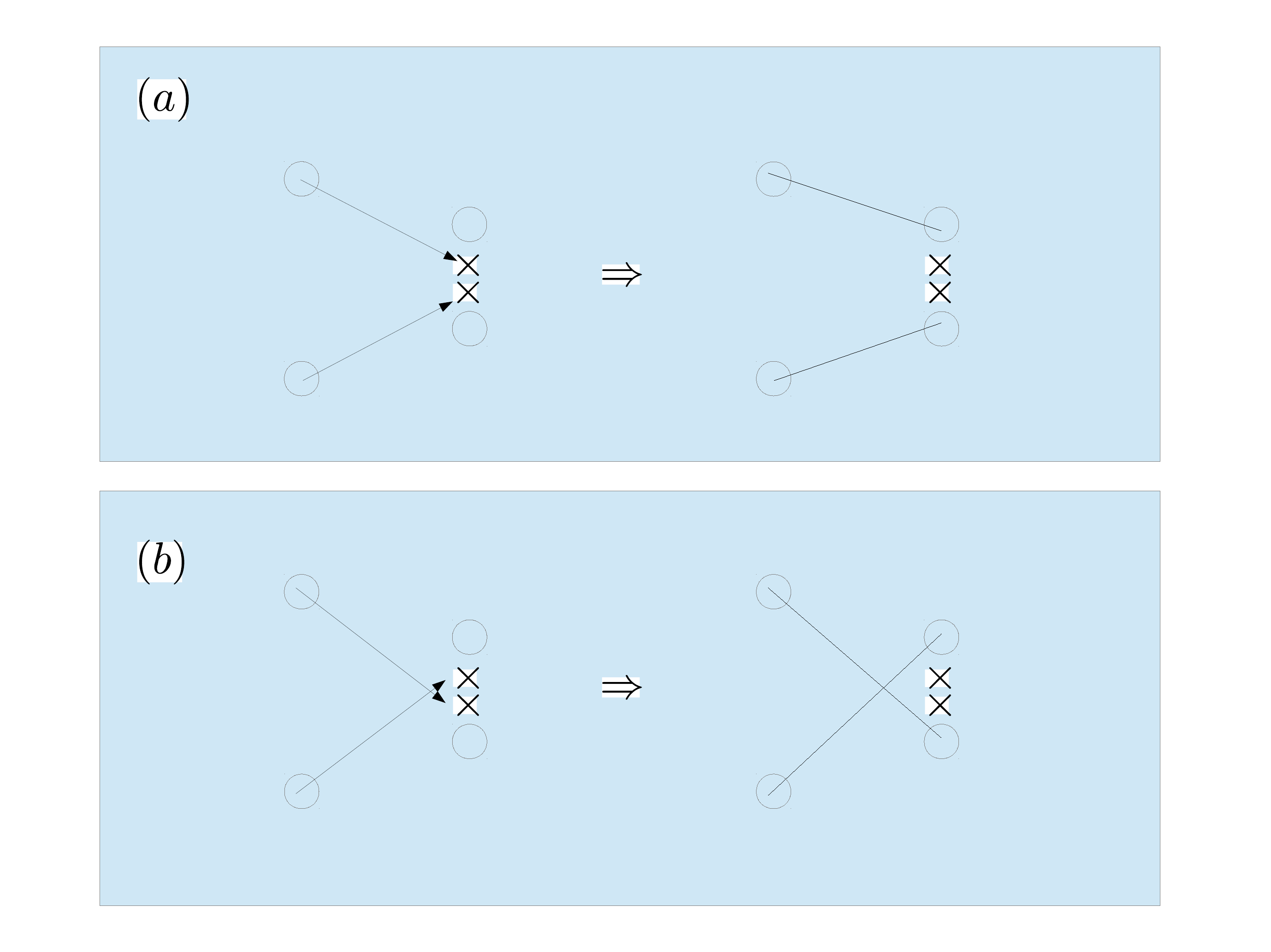}
\caption{
A schematic to demonstrate how the perturbed eigenvalues (crosses) obtained from the
perturbation theory are used to establish
two mode connectivities of actual eigenvalues (circles) from $\qvec_{i}$ to $\qvec_{i+1}$.
The actual eigenvalues are deduced from the diagonalizations of
dynamical matrices at $\qvec_{i}$ and $\qvec_{i+1}$. 
The scenarios of (a) noncrossing and (b) crossing of bands show the importance of 
the accuracy of the predicted eigenvalues for establishing the correct mode connectivities.
}
\label{fig:2}
\end{figure}

From the ordering deduced from the sorted perturbed eigenvalues $\Omega^2_{n,\qvec_{i+1}}$ and the ordering of the sorted actual eigenvalues $\omega^2_{m,\qvec_{i+1}}$ which we obtain from diagonalizing the dynamical matrix at $\qvec_{i+1}$, we establish a very reasonable mode connectivity from $\qvec_i$ to $\qvec_{i+1}$ by simply connecting the $\omega^2_{n,\qvec_{i}}$ and $\omega^2_{m,\qvec_{i+1}}$ guided by $\Omega^2_{n,\qvec_{i+1}}$, as can be seen in \Fig~\ref{fig:2}. When the connectivities between $\qvec_{i}$ and $\qvec_{i+1}$, we can use the
same perturbative approach to establish  connectivities between $\qvec_{i+1}$ and $\qvec_{i+2}$, etc.

We note that the perturbative approach for local connectivity may fail at $\qvec_i$ due to the presence of degenerate or near-degenerate modes for the following reason.
Suppose the correct scenario of the connectivities of the two modes is to cross as shown in \Fig~\ref{fig:2}(b). 
This switching has to be facilitated by the switching of the order of two predicted eigenvalues (as shown on the left of \Fig~\ref{fig:2}(b)). 
However, if these two predicted eigenvalues fail to switch due to reasons such as inaccuracies in the dynamical matrices, too large a $V_i$ term, or the predicted eigenvalues lying extremely close to one another as in the case of degenerate modes where it leads to an incorrect ordering of predicted eigenvalues as shown on the left of \Fig~\ref{fig:2}(a), then it will lead to an incorrect noncrossing scenario.  
This means that the perturbed eigenvalues must be correctly ordered if they are to give the correct mode connectivities.

\subsection{Global mode connectivity}
\label{subsec:GlobalConnectivity}

To overcome the shortcoming of the perturbative approach, we do not use it to connect the phonon modes for all $\qvec$ points between $\qvec_b$ and $\qvec_e$.
Instead, we only apply the perturbative approach in the first stage to connect the modes for the first $s$ $\qvec$ points, i.e., from $\qvec_0$, $\qvec_1$, $\cdots$, to $\qvec_{s-1}$, in order to construct the heads of the phonon branches, before switching in the second stage to an approach that is based on least-squares fits and exploits the smooth `global' structure of the phonon bands.
Beginning from $\qvec_{s}$, for each band index $n$, we use the previous $r$ (obviously $r \le s$) values of $\omega^2_{n,\qvec_{i}}$, for $i = s-1, s-2, \cdots, s-r$, to form a quadratic fit for each $n$ and predict the eigenvalue at $\qvec_{s}$.  
The subroutine for the least-squares fit of a polynomial is implemented according to Ref.~\cite{Anton1994-book}.
These eigenvalues $\Omega^2_{n,\qvec_s}$ predicted from the least-squares fit are sorted and compared with the sorted exact eigenvalues $\omega^2_{m,\qvec_s}$ to establish the mode connectivities from $\qvec_{s-1}$ to $\qvec_{s}$.  
To construct the remainder of the phonon path, we repeat the same fitting approach stepwise to connect the modes from $\qvec_{s}$ to $\qvec_{s+1}$, from $\qvec_{s+1}$ to $\qvec_{s+2}$, and so on until we finally connect all the modes from $\qvec_{p-1} $ to $\qvec_{p}$ for every band. 
This completes the first pass of mode connectivities from $\qvec_0$ to $\qvec_{p}$.

To eliminate the inappropriate connectivities that may have occurred during the first stage with the perturbative approach, we initiate a second pass of mode connectivities from $\qvec_{p}$ to $\qvec_0$, where the last $r$ eigenvalues for modes are taken from $\qvec_{p}$, $\qvec_{p-1}$, $\cdots$, $\qvec_{p-r+1}$ for fitting purposes, and establish connectivities between $\qvec_{p-r+1}$ and $\qvec_{p-r}$. 
Thereafter, the fitting approach is used to move stepwise from $\qvec_{p-r}$ to $\qvec_{p-r-1}$ and so on until we finally connect $\qvec_{1}$ to $\qvec_{0}$. 
We note here that since all $\qvec$ paths are independent of one another as far as the band indices are concerned, we have to apply the hybrid method afresh at the beginning of each path.

We now explain why the fitting approach handles the near or exact degenerate modes well. 
This is largely due to the fact that when a value is taken from the list of very close eigenvalues to be part of an array of $r$ values for a quadratic fit, the predicted eigenvalue is likely to be very insensitive to any value chosen from the set of nearly degenerate eigenvalues since all values in the set are very close to one another.
We also point out that our hybrid method is applicable to almost any phonon code (DFPT based or small-displacement based) since it is easy to generate the input dynamical matrices at any $\qvec$ points with the  knowledge of interatomic force constants.

\section{Results}
\label{sec:results}

\begin{figure}
\centering\includegraphics[width=8.2cm,clip]{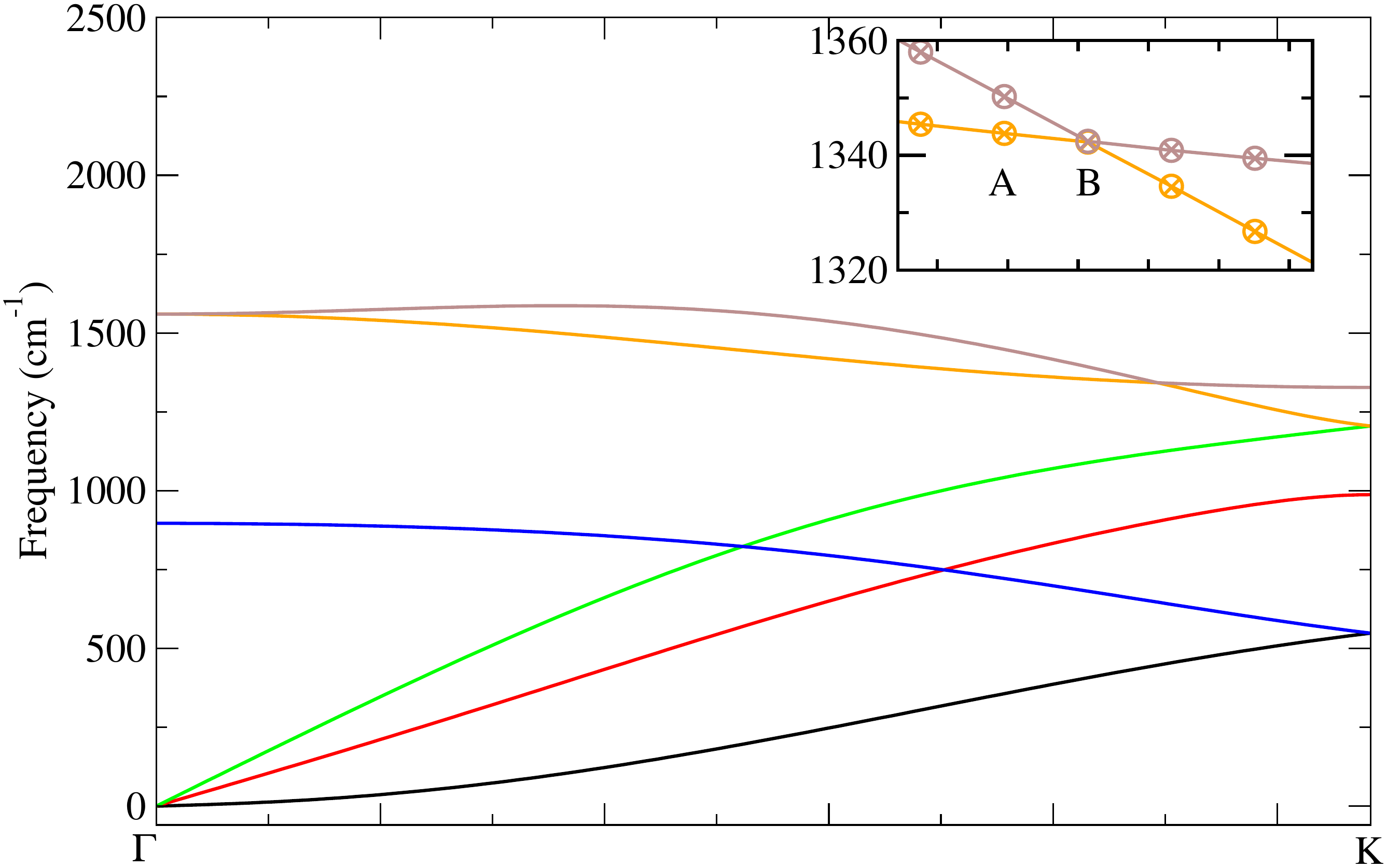}
\caption{
Incorrect phonon mode connectivities are observed at the upper two
branches for graphene when only the perturbative approach (i.e.,
without the fitting approach) is used.  Even though the inset shows
very good agreement between the actual (circles) and predicted
eigenvalues (crosses), the perturbative approach (with $\tau =
0.5$~\invcm{}) alone fails to give a correct order of eigenvalues at
$B$ resulting a kink.  Interestingly, the perturbative approach gives
correct connectivities at two other crossing points at about
$750$~\invcm{}.  The selected $\VEC{q}$ points (in $\VEC{b}_1$,
$\VEC{b}_2$, and $\VEC{b}_3$) are $\Gamma=[0,0,0]$, $K =
[\frac{1}{3},\frac{1}{3},0]$.  $|\delta \qvec| = 0.0149$~\invA{}.
}
\label{fig:grapheneGK}
\end{figure}

\begin{figure}
\centering\includegraphics[width=8.2cm,clip]{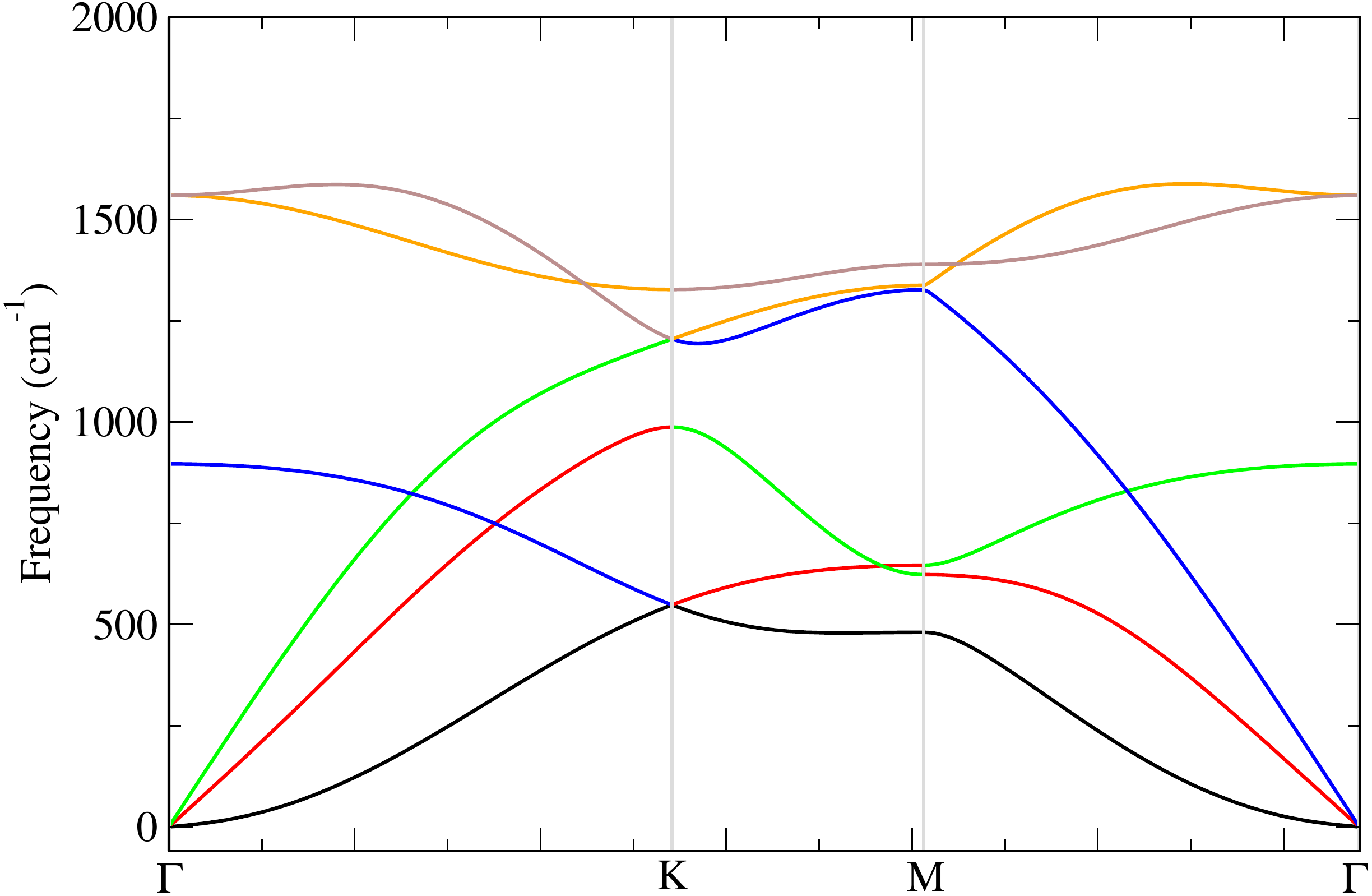}
\caption{
Phonon mode connectivities of graphene using the hybrid method with
$\tau=0.5$~\invcm{}, $s=4$, $r=4$, and quadratic fits.  Notice that
correct connectivities (c.f. \Fig~\ref{fig:grapheneGK}) are
restored at the top two bands for the $\Gamma\smd{}K$ path.  The selected $\VEC{q}$ points (in
$\VEC{b}_1$, $\VEC{b}_2$, and $\VEC{b}_3$) are $\Gamma=[0,0,0]$, $K =
[\frac{1}{3},\frac{1}{3},0]$, and $M= [0,\frac{1}{2},0]$.  $|\delta
\qvec| = 0.0149$, $0.0147$, and $0.0147$~\invA{} for the $\Gamma\smd{}K$,
$K\smd{}M$, and $M\smd{}\Gamma$ paths, respectively.
}
\label{fig:graphene}
\end{figure}

To assess the efficacy of the proposed hybrid method, we carry out density-functional theory (DFT) calculations within the local density approximation as implemented in the Vienna Ab initio Simulation
Package (VASP)\cite{Kresse96v6}, with projected augmented-wave (PAW) pseudopotentials.  
The phonon calculations are performed using a small-displacement method\cite{Gan21v259,Gan10v49}.  
The first system is a 2D graphene sheet with lattice constant $a=2.462$~\AA{} and a vacuum height of $12$~\AA{}.  
A $4\times 4 \times 1$ supercell is used. 
The cutoff energy for the plane-wave basis set is $500$~eV. 
A $k$ mesh of $6 \times 6 \times 1$ is used for electronic relaxation.
Figure~\ref{fig:grapheneGK} shows the mode connectivities of graphene along the $\Gamma\smd{}K$ path with only the perturbative approach (i.e., we do not use the fitting approach).  
It is seen that the upper two bands have wrong connectivities near $1340$~\invcm{}.  
The inset in \Fig~\ref{fig:grapheneGK} shows why the perturbation approach fails to predict the correct ordering of eigenvalues at the degenerate point. 
Here, the eigenvalues in circles are obtained from exact diagonalization and the eigenvalues in crosses are obtained from perturbation theory. 
The perturbative approach connects well the modes from $\Gamma$ to a $\qvec$ point corresponding to $A$. 
Because of the wrong ordering for the predicted eigenvalues at a $\qvec$ point corresponding to $B$, a wrong assignment of modes occurs that results in visually detectable kinks in \Fig~\ref{fig:grapheneGK}.
However, when the hybrid approach is used, it is found that the fitting approach is able to overcome the problem highlighted in Fig.~\ref{fig:grapheneGK} and gives correct connectivities
in the $\Gamma\smd{}K$ path shown in \Fig~\ref{fig:graphene}. 
It is also observed that other band crossings along the $K\smd{}M$ (one intersection point) and $M\smd{}\Gamma$ (two intersection points) paths
are correctly produced.

\begin{figure}
\centering\includegraphics[width=8.2cm,clip]{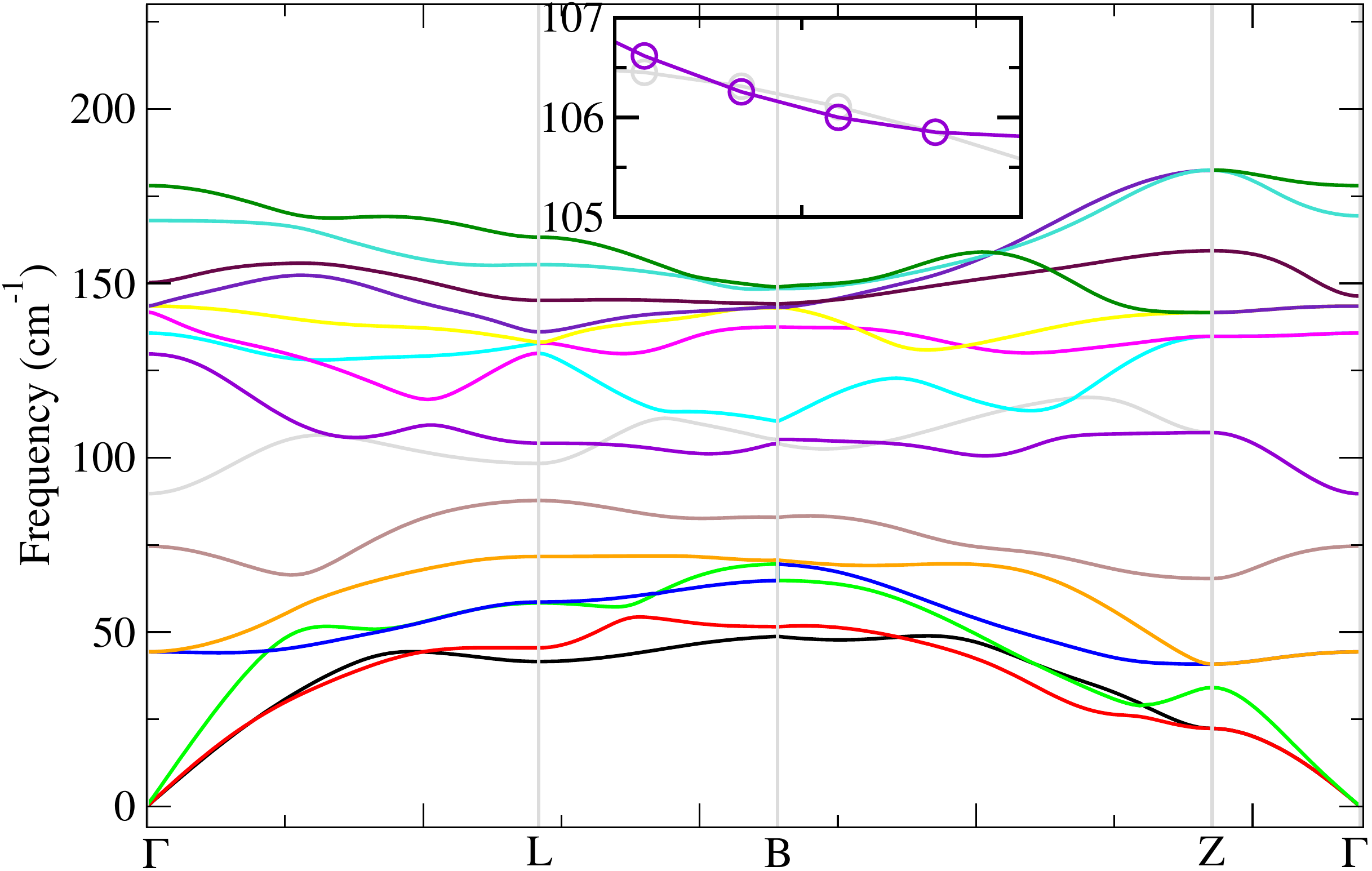}
\caption{
Phonon mode connectivities of \bise{} obtained using the hybrid method
with $\tau=0.5$~\invcm{}, $s=4$, $r=4$, and quadratic fits.  The
insets shows delicate connectivities along the $\Gamma\smd{}L$ path
near $106$~\invcm{}.  The selected $\VEC{q}$ points (in $\VEC{b}_1$,
$\VEC{b}_2$, and $\VEC{b}_3$) are $\Gamma=[0,0,0]$, $L= [\half,0,0]$,
$B=[\eta,\half,1-\eta]$, and $Z=[\half,\half,\half]$, where $\eta =
(1+4 \cos \alpha_r)/ (2+ 4\cos\alpha_r)$.  $|\delta \qvec| = 0.0148$,
$0.0143$, $0.0150$, and $0.0140$~\invA{} for the $\Gamma\smd{}L$,
$L\smd{}B$, $B\smd{}Z$, and $Z\smd{}\Gamma$ paths, respectively.  The
effect of longitudinal optical (LO) and transverse optical (TO)
splitting\cite{Liu14v16} has been taken account.
}
\label{fig:bi2se3}
\end{figure}

Figure~\ref{fig:bi2se3} shows the mode connectivities for the trigonal \bise{} with $a_r=9.621$~\AA, $\alpha_r=24.64^\circ$, which corresponds to a conventional hexagonal cell of $a= 4.105 $ and $c=27.973$~\AA{}. 
The supercell is $4 \times 4 \times 1$ of the conventional hexagonal unit cell. 
The cutoff energy for the plane-wave basis set is $423.2$~eV. 
A $k$ mesh of $4 \times 4 \times 2$ is used for electronic relaxation.  
The intricate band crossings at about $106$~\invcm{}, near the midpoint of the $\Gamma\smd{}L$ path, appear to be very well captured by the hybrid method as shown in the inset of \Fig~\ref{fig:bi2se3}. 
We find that the hybrid method is robust enough to handle the three nearly degenerate points of about $160$~\invcm{} 
in the $B\smd{}Z$ path.

\begin{figure}
\centering\includegraphics[width=8.2cm,clip]{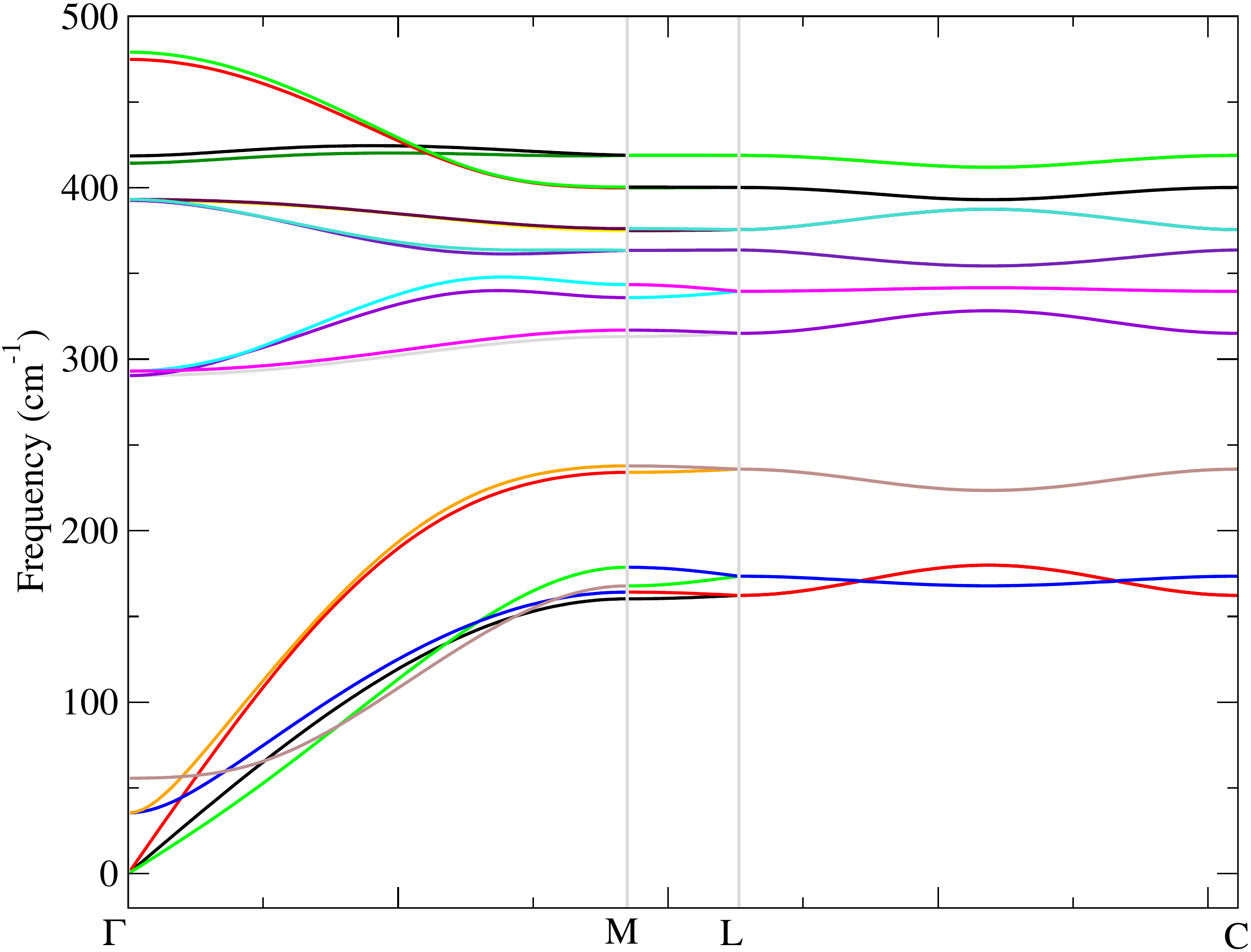}
\caption{
Phonon mode connectivities of \mos{} obtained using the hybrid method
with $\tau=0.5$~\invcm{}, $s= 4$, $r=4$, and quadratic fits.  The
selected $\VEC{q}$ points (in $\VEC{b}_1$, $\VEC{b}_2$, and
$\VEC{b}_3$) are $\Gamma=[0,0,0]$, $M= [0,\half,0]$,
$L=[0,\half,\half]$, and $C=[\half,\half,\half]$.  $|\delta \qvec| =
0.0171$, $0.0162$, and $0.0171$~\invA{} for the $\Gamma\smd{}M$,
$M\smd{}L$, and $L\smd{}C$ paths, respectively.  The effect of
longitudinal optical (LO) and transverse optical (TO)
splitting\cite{Liu14v16} has been taken account.
}
\label{fig:mos2}
\end{figure}

Figure~\ref{fig:mos2} shows the mode connectivities of the hexagonal \mos{} with $a=3.123$ and $c = 12.087$~\AA{}.  
The supercell size is $3\times 3\times 2$. 
The cutoff energy for the plane-wave basis set is $700$~eV. A $k$ mesh of $2\times 2\times 1$ is used for the electronic
relaxation. We observe very good connectivities along all paths. For example,
along the $\Gamma\smd{}M$ path, the somewhat dispersive 
bands below $180$~\invcm{} give rise to many band crossings all of which the hybrid method is capable of resolving. 
For the high-frequency modes of about $420$~\invcm{}, four crossing points in a small region are shown to be 
described correctly. Note that all bands are doubly degenerate in the $L\smd{}C$ path, which pose no difficulty for the hybrid method.


\section{Conclusion}
\label{sec:conclusion}

In this paper we have proposed a two-stage hybrid method that uses local and global information for phonon mode connectivity. 
In the first stage of the algorithm, we exploit the local continuity of the eigenvalues within each band to construct part of the phonon dispersion path. 
We connect the actual eigenvalues at $\qvec_{i}$ to actual eigenvalues at $\qvec_{i+1}$ through the use of first-order degenerate perturbation theory in which the perturbed eigenvalues at $\qvec_{i+1}$ 
are predicted from the eigenvectors at $\qvec_{i}$. 
In the second stage, we use a global polynomial fit of the partially constructed phonon dispersion path, which is based on least-squares fits, to predict the eigenvalues and then connect the eigenvalues. 
This approach avoids local errors associated with degenerate or near-degenerate eigenvalues at or near the crossing points.

We find that a moderately fine $\qvec$ mesh is sufficient for the application of our proposed hybrid method, in contrast with a very fine $\qvec$ mesh used
in the commonly adopted eigenvector similarity method.  
We note that our method is applicable to any phonon code (density-functional perturbation theory based or small-displacement based) since the phonon code generates the input dynamical matrices at any $\qvec${} point to our hybrid method.  
The robustness of our method has been demonstrated for graphene, \bise{}, and \mos{}.  
We expect the use of our hybrid method could be be extended to handle mode connectivities in electronic band structures, or photonic modes in photonic crystals.

\begin{acknowledgments}
We gratefully acknowledge support from the Science and Engineering Research Council through two grants (a) 152-70-00017, and (b) RIE2020 Advanced Manufacturing and Engineering (AME) Programmatic Grant No
A1898b0043.  
We thank the National Supercomputing Center, Singapore (NSCC) and A*STAR Computational Resource Center, Singapore (ACRC) for computing resources.
\end{acknowledgments}

\end{document}